# Mixed RF/FSO Cooperative Relaying Systems With Co-Channel Interference

Elyes Balti, *Student Member, IEEE*, and Mohsen Guizani, *Fellow, IEEE*

*Abstract*—In this paper, we provide a global framework analysis of a dual-hop mixed radio frequency (RF)/free space optical (FSO) system with multiple branches/relays wherein the first and second hops, respectively, consist of RF and FSO channels. To cover various cases of fading, we propose generalized channels' models for RF and FSO links that follow the Nakagami-m and the double generalized gamma distributions, respectively. Moreover, we suggest channel state information (CSI)-assisted relaying or variable relaying gain based amplify-and-forward amplification. Partial relay selection with outdated CSI is assumed as a relay selection protocol based on the knowledge of the RF CSI. In order to derive the end-to-end signal-to-interference-plus-noise ratio statistics, such as the cumulative distribution function, the probability density function, the higher order moments, the amount of fading and the moment generating function, the numerical values of the fading severity parameters are only valid for integer values. Based on these statistics, we derive closed-forms of the outage probability, the bit error probability, the ergodic capacity, and the outage capacity in terms of Meijer-G, univariate, bivariate, and trivariate Fox-H functions. Capitalizing on these expressions, we derive the asymptotic high SNR to unpack valuable engineering insights of the system performance. Monte Carlo simulation is used to confirm the analytical expressions.

## I. Introduction

WITH the extremely high demand for the bandwidth, Radio Frequency (RF) technology, which is the second mostly used for the backhaul networking after the copper lines and represents 6% of the total used transport media in the US [1], becomes unable to support the big data flows of the large number of users since the spectrum is limited and the access license is very costly. Moreover, shared utilization of the bandwith between the primary and secondary users based systems reaches their bottlenecks since the last ones still suffer from the spectrum scarcity. Therefore, current RF systems cannot support the high performance requirements of the fifth generation (5G) standards and future mobile broadband networks such as 3GPP LTE-advanced, IEEE 802.16m, and IEEE 802.16j. To overtake this critical situation, recent research attempts have proposed the usage of the optical fibers (OF) as a way to reduce the congestion of the backhaul networks. Unlike microwave and mmWave (from 6 to 300 GHz) channels, OF provides not only high rate communications over long distance, e.g., 155.52 Mbit/s for STM-1, 622 Mbit/s for STM-4, 2.4 Gbit/s for STM-16, and 9.9 Gbit/s for STM-64, but also it is immune against the interference problems and low coverage. In addition, since they are very expensive to be installed and need important investment [2], the total usage of OF for backhauling in US is below 4%. The main drawback of OF is that they cannot be deployed in some restricted areas and applications. In this case, OF cannot be reliable for ultra dense networks wherein a considerable deployment of OF is required to serve the enormous demand of microcell, picocell, and femtocell, etc.

To address this shortcoming, Free Space Optical (FSO) communications are recently proposed as an alternative or complementary to RF and OF solutions due its flexibility, free spectrum access license, immunity to interference, high security level, power efficiency, cost effectiveness, no installation restriction and most importantly it is a way to densify the cellular networks [3]–[10]. These features make the FSO links' capacities 25 fold more efficient than RF technology and essentially they are a cost-efficient solution compared to OF [11]. Because of these advantages, FSO becomes a promising solution for the last mile problem to bridge the bandwidth gap between the end-users and the OF backbone network. Based on the aforementioned points, FSO has been used both in academia and industry such as enterprise/campus connectivities, video surveillance, redundant links, disaster recovery, security, and broadcasting [7].

### A. Motivation

Although FSO links have recently gained enormous attention due to the aforementioned advantages, they are extremely sensitive to atmospheric weather conditions, e.g., rain, fog, and snow [7], [12]. Standard values of the atmospheric path loss relative to the weather conditions are detailed and given by [13], [14]. In addition, FSO links experience the turbulence-induced fading of atmoshperic eddies, which can be modeled by Log-Normal distribution. This model is widely used in particular for weak turbulence where it provides a good fit to the experimental data [13], [15]. As the turbulence becomes moderate, Log-Normal model deviates from the standard data and cannot fully characterize the turbulence conditions. To address this limitation, Gamma-Gamma ($G^2$) has been recently proposed to cover wider range of turbulences from weak to moderate conditions [16], [17]. As the turbulences become severe, $G^2$ also loses its accuracy in describing the experimental data in particualr at the tail data region. To overtake this shortcoming, the so-called Double Generalized Gamma (DGG) is recently introduced by Kashani *et. al* [18], which not only reflects various ranges of turbulence but also provides the best curve fitting to the data mainly at the tail. In addition to the path loss and turbulence fading, FSO links are also subject to the pointing error, which is originated from the misalignment between the laser-emitting transmitter and the photodetector. Various factors such as seismic activities, building sways results in the aforementioned misalignment. To characterize this degradation, Uysal *et. al* [19] discuss various models for the radial displacement of the pointing error for a Gaussian laser beam. The most generalized model that covers various special cases is the so-called Beckmann pointing error model. Based on this model, related work have adopted the following choices to model the radial displacement such as Rician [20], Hoyt [21], NonZero-Mean and Zero-Mean Single-Sided Gaussian [22] but the most widely used is Rayleigh [17], [23], [24] for a reason of simplicity.

Owing to the relative atmoshperic constraints imposed on the FSO links, the best way is to propose an adpative solution combining both RF and FSO links within the same system. In clear weather conditions, only FSO links are activated to transfer the data, however as the atmospheric conditions becomes harsh, the data is sent over RF links [12]. In this context, mixed RF/FSO systems have recently attracted various research fields since it combines the reliability of RF communications in severe conditions and both cost-efficient and capacities of FSO links [25]–[28].

In the same context, mixed RF/FSO can operate over long distances to serve more cells in farther areas such as mountains and forests, etc. In reality, the coverage varies from area to another and depends mainly on the distance between the centralized core network and the cells or the base stations. Over long distances, the power transmission may suffer from decaying and attenuation due to various factors such as ohmic resistance, thermal and radiation resistance. A practical solution to address this mitigation is to deploy relays along the intermediate paths to amplify, denoise and improve the signal quality. In this case, the coverage and the network scalability substantially enhances and the system becomes reliable over long distances. Because of the mentioned benefits, cooperative relaying communication becomes the promising solution for mixed RF/FSO system and the corner-stone since it provides possible coverage extention, uniform quality of service (QoS), spatial diversity gain and hotspot throughput improvement [29]–[32]. Furthermore, there are various relaying modes discussed in the literature but the most widely used are Amplify-and-Forward (AF) with fixed and variable relaying gain [33]–[36], Decode-and-Forward (DF) [37], [38], Quantize-and-Encode (QE) [39] and Quantize-and-Forward (QF) [40].

In practice, cellular systems require a large number of relays wherein they could be deployed in series or parallel depending on the topology and the network configuration. Based on the network topology, previous attempts have proposed different protocols to manage the relaying functioning such as all-active relaying and selective relaying. All-active relaying consists of simultaneous parallel transmissions to all relays [41] while selective relaying is based on selecting one relay among the set following predefined rules. Various relay selection protocols have been developed in the literature such as opportunistic relay selection [42], partial relay selection (PRS) [42], distributed switch and stay, max-select protocol, and all active relaying [41]. Although all-active relaying seems to be more efficient than selective relaying, it is restricted in parallel FSO communications because of the photodetector synchronization problem. As we mentioned, selective relaying protocols are based on various rules and parameters but the most important factor is the Channel State Information (CSI). The source (S) and the Destination (D) do not always have a full knowledge of the channels' gains and instead they estimate the channels' coefficients based on the CSI feedback delivered by the relays. For low time-varying channels, the channel coherence time is enough significant that $S$ could easily retrieve the channels' coefficients and then has a perfect channel estimation based on the CSI feedback. However for rapid time-varying channels, the channels' coefficients are rapidly changing and given that the CSI feedback is slowly propagating, $S$ will be unable to perfectly estimate the gains. In this case, it is straightforward to assume an outdated CSI rather than perfect estimation [17], [43].

*B. Related Work*

Enormous work dealing with mixed RF/FSO relaying systems have been proposed in the literature. Soleimani-Nasab and Uysal [23] considered a dual-hop mixed RF/FSO system with co-channel interference and line of sight where the RF and FSO channels experience Nakagami-m and DGG, respectively. Rayleigh is a special case of Nakagami-m as it was considered in [40] and [42], while the Rician fading, assumed in [62] and [41], can be approximated to have Nakagami-m model. The mentioned work mostly assumed $G^2$ as an FSO fading model except in [36] where Málaga distribution was assumed. Furthermore, Al-Quwaiee *et. al* in [48] derived the statistics of the end-to-end Signal-to-Noise Ratio (SNR) and based on that they also provided closed-forms of the outage probability, the bit error probability and the ergodic capacity. Besides, Yang *et. al* in [49] derived the same performance achieved by [50] but they assumed transmit diversity at the source and selection combining at the receiver. Further work [51], [52] proposed hybrid RF/FSO systems with multiple relays and outdated CSI assuming an aggregate model of hardware impairments introduced at $S$ and the relays.

*C. Contribution*

In this paper, we propose a dual-hop mixed RF/FSO system with multiple relays where RF channels experience Nakagami-m fading and FSO links are subject to DGG fading encompassing the turbulence-induced fading, atmospheric path loss, and pointing error. We also consider the co-channel interference, which is detrimental to RF links. Besides, the relays employ AF with CSI-assisted relaying and we consider partial relay selection with outdated CSI based on the RF channels information. Furthermore, the photodetector can detect the signal following either the coherent/heterodyne mode or the Intensity Modulation and Direct Detection (IM/DD). In addition, Subcarrier Intensity Modulation (SIM) is implemented into the relays to modulate the intensity of the FSO carriers. Various binary modulation schemes are assumed to validate the error performance of the proposed system. To the best of our knowledge, our work is a generalization of the existing related work. The analysis of this paper follows these steps:

1) Present a detailed analysis of the system and channels' models.
2) Provide the Cumulative Distribution Function (CDF) and the Probability Density Function (PDF) of the RF and FSO channels.
3) Derive the statistics of the end-to-end Signal-to-Interference-plus-Noise Ratio (SINR) such as the CDF, PDF, high order moment, amount of fading, and the Moment Generating Function (MGF).
4) Based on the aforementioned statistics, novel closed-forms as well as high SNR asymptotes of the outage probability, the bit error probability, the ergodic capacity, and the outage rate are derived.
5) Capitalizing on the asymptotic high SNR, engineering insight into the system gains such as the diversity gain is derived.

*D. Structure*

This paper is organized as follows: Section II describes the system model while Section III provides the statistics of the overall SINR. Performance analysis are provided in Section IV while numerical results following their discussions are presented in Section V. Concluding remarks are given in Section VI.

*E. Notation*

For the sake of organization, we provide some useful notations to avoid the repetition. $f_\mathrm{h}(\cdot)$ and $F_\mathrm{h}(\cdot)$ denote the PDF and CDF of the random variable $h$, respectively. Gamma distribution with parameters $\alpha$ and $\beta$ is denoted by $\mathcal{G}(\alpha,\beta)$ while the Generalized Gamma distribution with parameters $\alpha$, $\beta$ and $\gamma$ is given by $\mathcal{GG}(\alpha,\beta,\gamma)$. In addition, the Gaussian distribution of parameter $\mu$, $\sigma^2$ is denoted by $\mathcal{N}(\mu,\sigma^2)$. The operator $\mathbb{E}[\cdot]$ stands for the expectation while $\Pr(\cdot)$ denotes the probability measure. The symbol $\backsim$ stands for "distributed as".

## II. SYSTEM AND CSIS MODELS

*A. System Model*

The proposed system consists of $M$ parallel relays wirelessly connected to $S$ and $D$. Partial relay selection based on the knowledge of the RF channels is assumed to select one relay among the set. This protocol states that for a given communication, $S$ periodically receives CSI feedback ($\gamma_{1(n)}$ for $n = 1, \ldots M$) from the relays, sorts them in an increasing order of magnitude and then select the branch/relay with the highest CSI. Hence, partial relay selection consists of selecting the $m$-th worst or $(M - m)$-th best relay $R_{(m)}$. Once $S$ receives the feedback, a processing time is required for resources allocation, prescheduling, etc. Given that the channels are time-varying, the received CSIs rapidly change after the processing time and hence the selection is achieved based on an outdated CSI. To model the relation between the updated and outdated CSIs, we define the time correlation coefficient $\rho$ as follows

$$\gamma_{1(\mathrm{m})} = \sqrt{\rho}\,\hat{\gamma}_{1(\mathrm{m})} + \sqrt{1-\rho}\,w, \tag{1}$$

where $\gamma_{1(\mathrm{m})}$ is the instantaneous CSI of the $m$th RF channel, $w \backsim \mathcal{N}(0,\sigma^2_{\gamma_{1(\mathrm{m})}})$, $\sigma^2_{\gamma_{1(\mathrm{m})}}$ is the variance of the $m$-th channel/CSI $\gamma_{1(\mathrm{m})}$. Note that the subscript of $\gamma_{1(\mathrm{m})}$ contains "1(m)" to indicate the $m$th channel of the first hop. The same notation is adopted for the channels of the second hop as $\gamma_{2(\mathrm{m})}$. The correlation coefficient $\rho$ is given by the Jakes' autocorrelation model as follows [53]

$$\rho = J_0(2\pi f_d T_d), \tag{2}$$

where $J_\nu(\cdot)$ is the $\nu$-th order Bessel function of the first kind, $T_d$ is the time delay between the current and the delayed CSI versions, and $f_d$ is the maximum Doppler frequency of the channels.

The received RF signal at the $m$-th relay is given by:

$$y_{1(\mathrm{m})} = h_{1(\mathrm{m})}x + \sum_{n=1}^{M_R} f_n d_n + \nu_{\mathrm{SR}}, \tag{3}$$

where $h_{1(\mathrm{m})}$ is the $m$-th channel gain of the first hop, $x$ is the information signal, $d_n$ is the modulation symbol of the $n$-th interferer with an average power $\mathbb{E}\left[|d_n|^2\right] = P_{R_n}$, $M_R$ is the number of interferers, $f_n$ is the fading between the $n$-th interferer and the selected relay and $\nu_{\mathrm{SR}}$ is the additive white Gaussian noise (AWGN) of the RF channels with variance $\sigma_0^2$. The received signal at $D$ can be expressed as follows

$$y_{2(\mathrm{m})} = (\eta I_{2(\mathrm{m})})^{\frac{r}{2}} Gh_{\mathrm{SR}}x + (\eta I_{2(\mathrm{m})})^{\frac{r}{2}} G\sum_{n=1}^{M_R} f_n d_n \\ + (\eta I_{2(\mathrm{m})})^{\frac{r}{2}} G\nu_{\mathrm{SR}} + \nu_{\mathrm{RD}}, \tag{4}$$

where $\eta$ is the electrical-to-optical conversion coefficient, $G$ is the relaying gain, $I_{2(\mathrm{m})}$ is the $m$-th FSO channel, $\nu_{\mathrm{RD}}$ is the AWGN of the FSO channels with variance $\sigma_0^2$, $r = 1$ and $r = 2$ represent the heterodyne detection and IM/DD, respectively. An illustrative system model is given by Fig. 1, where the mmWave channels connect the different mobile users to the base stations. The FSO links play the role of back-hauling to connect the various networks such as the ISP (Internet Service Provider), mobile network, and enterprise network to the main data centers.

*B. CSIs Model*

Since the outdated RF CSI $\backsim \mathcal{G}(m_{\mathrm{SR}},\Omega_{\mathrm{SR}}/m_{\mathrm{SR}})$, the PDF and CDF of the instantaneous SNR are expressed as

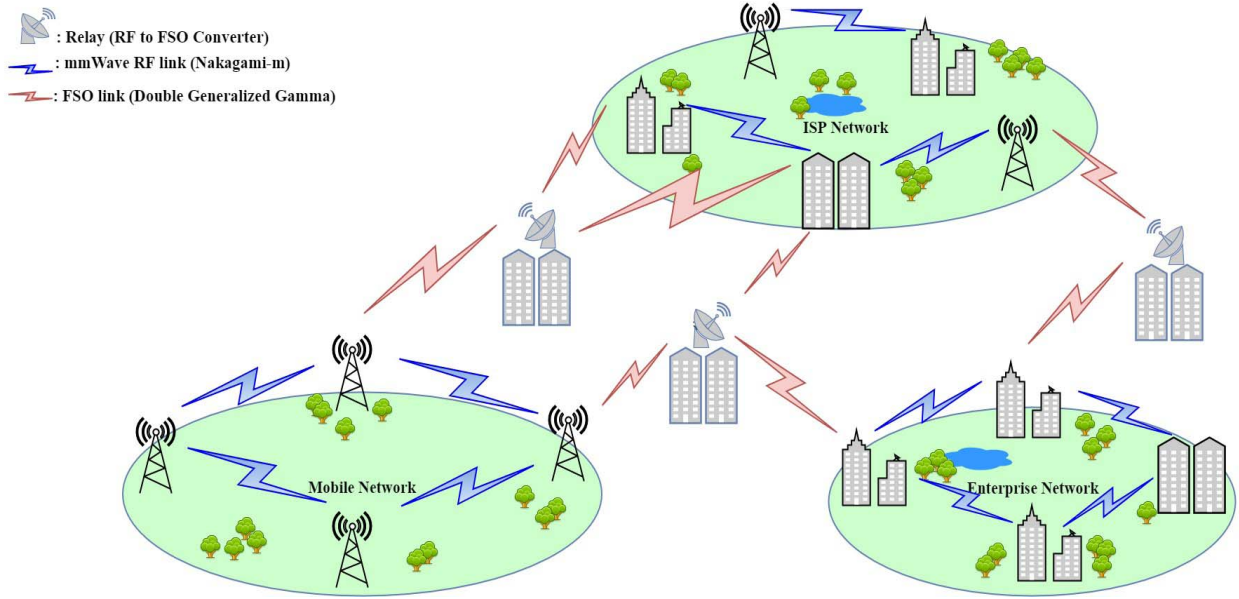

Fig. 1. Mixed RF/FSO relaying system.

follows

$$f_{\hat{\gamma}_{1(m)}}(\gamma) = \frac{\alpha_{SR}^{m_{SR}} \gamma^{m_{SR}-1}}{\Gamma(m_{SR})} e^{-\alpha_{SR}\gamma}, \quad (5)$$

$$F_{\hat{\gamma}_{1(m)}}(\gamma) = 1 - \frac{\Gamma(m_{SR}, \alpha_{SR}\gamma)}{\Gamma(m_{SR})}, \quad (6)$$

where $\alpha_{SR} = \frac{m_{SR}}{\Omega_{SR}}$.

To simplify the mathematical derivations, we assume that $m_{SR}$ is an integer and hence the CDF can be reformulated as follows

$$F_{\hat{\gamma}_{1(m)}}(\gamma) = 1 - e^{-\frac{m_{SR}\gamma}{\overline{\gamma}_{SR}}} \sum_{n=1}^{m_{SR}-1} \frac{1}{n!} \left(\frac{m_{SR}\gamma}{\overline{\gamma}_{SR}}\right)^n, \quad (7)$$

where $\overline{\gamma}_{SR}$ is the average SNR of the RF link.

The outdated and current instantaneous SNRs are jointly Nakagami-m distributed with the joint PDF expressed as follows

$$f_{\gamma_{1(m)}, \hat{\gamma}_{1(m)}}(x,y) = \left(\frac{m_{SR}}{\overline{\gamma}_{SR}}\right)^{m_{SR}+1} \frac{\left(\frac{xy}{\rho}\right)^{\frac{m_{SR}-1}{2}}}{(1-\rho)\Gamma(m_{SR})} \\ \times e^{-\frac{m_{SR}}{\overline{\gamma}_{SR}}\left(\frac{x+y}{1-\rho}\right)} I_{m_{SR}-1}\left(\frac{2m_{SR}\sqrt{\rho xy}}{\overline{\gamma}_{SR}(1-\rho)}\right), \quad (8)$$

After some mathematical manipulations, the PDF of the current instantaneous SNR is given by [54, eq. (3.11)]

$$f_{\gamma_{1(m)}}(\gamma) = \frac{m_{SR}}{\Gamma(m_{SR})} \binom{M}{m_{SR}} \sum_{n=0}^{m_{SR}-1} \sum_{i=0}^{j(m_{SR}-1)} \sum_{v=0}^{i} \binom{m_{SR}-1}{n} \\ \times \binom{i}{v} \left(\frac{m_{SR}}{\overline{\gamma}_{SR}}\right)^{m_{SR}+v} \Xi_{m_{SR}-1}^{i,j} \gamma^{m_{SR}+v-1} \\ \times \frac{(-1)^n \rho^v \Gamma(m_{SR}+i)}{(1+j(1-\rho))^{m_{SR}+v+i}(1-\rho)^{v-i}\Gamma(m_{SR}+v)} \\ \times \exp\left[-\frac{m_{SR}(j+1)\gamma}{(1+j(1-\rho))\overline{\gamma}_{SR}}\right], \quad (9)$$

where $I_\nu(\cdot)$ denotes the $\nu$-th order modified Bessel function of first kind and the coefficients $\Xi_m^{i,j}$ are defined and evaluated recursively as $\left(\sum_{i=0}^{m} \frac{x^i}{i!}\right)^j \triangleq \sum_{i=0}^{j(m-1)} \Xi_m^{i,j} x^i$, $\Xi_m^{i,j} \triangleq \sum_{n=n_1}^{n_2} \frac{\Xi_m^{n_1,j-1}}{(i-n_1)!} x^i$, $n_1 = \max(0, i - m_{SR})$, $n_2 = \min(i, (j-1)(m_{SR}-1))$ [55].

The instantaneous SNR of each interferer $\gamma_{R,k} \curvearrowright \mathcal{G}(m_{R,k}, 1/\beta_R)$ where $\beta_R \triangleq \frac{m_{R,k}\sigma_0^2}{\Omega_{R,k}P_{R_k}}$, $(m_{R,k}, \Omega_{R,k})$ are Nakagami-$m$ parameters between the $k$-th interferer and the relay. It has been shown in [56] that the sum of $L$ i.i.d Gamma random variables with shape parameter $\sigma$ and scale parameter $\alpha$ is a Gamma random variable with parameters $\sigma L$ and $\alpha$. The PDF of the total Interference-to-Noise Ratio (INR) $\gamma_R \triangleq \sum_{k=1}^{M_R} \gamma_{R,k}$ can be expressed as follows

$$f_{\gamma_R}(\gamma) = \frac{\beta_R^{m_R}}{\Gamma(m_R)} \gamma^{m_R-1} e^{-\beta_R \gamma}, \quad (10)$$

where $m_R \triangleq \sum_{k=1}^{M_R} m_{R,k}$.

The FSO fading encompasses the turbulence-induced fading ($I_a$), the atmospheric path loss ($I_l$) and the pointing errors ($I_p$). The $m$-th channel gain $I_{2(m)}$ can be written as follows

$$I_{2(m)} = I_a I_l I_p, \quad (11)$$

In Table I summarizes the parameters of the optical part.

Using the Beers-Lambert law, the path loss can be expressed as follows [17, eq. (12)]

$$I_l = \exp(-\sigma L), \quad (12)$$

The pointing error $I_p$ made by Jitter can be given as [13, eq. (9)]

$$I_p = A_0 \exp\left(-\frac{2R^2}{\omega_{Leq}^2}\right), \quad (13)$$

The atmospheric turbulence fading $I_a$ consists of small scale ($I_x$) and large scale ($I_y$) where $I_x \sim \mathcal{GG}(\alpha_1, m_1, \Omega_1)$ and $I_y \sim \mathcal{GG}(\alpha_2, m_2, \Omega_2)$, $m_1$ and $m_2$ are the shaping parameters defining the atmospheric turbulence fading. Moreover, $\alpha_1, \alpha_2, \Omega_1$, and $\Omega_2$ are defined using the variances of the small and large scale fluctuations from [18, eqs. (8.a), (8.b), (9), (10)]. Thereby, the PDF of the turbulence-induced fading $I_a$ can be given by [18, eq. (4)]

$$f_{I_a}(I_a) = \frac{\alpha_2 p^{m_2+\frac{1}{2}} q^{m_1-\frac{1}{2}} (2\pi)^{1-\frac{p+q}{2}}}{\Gamma(m_1)\Gamma(m_2) I_a} \times G_{p+q,0}^{0,p+q}\left(\frac{p^p q^q \Omega_1^q \Omega_2^p}{m_1^q m_2^p I_a^{\alpha_2 p}} \middle| \begin{array}{c} \Delta(q:1-m_1),\ \Delta(p:1-m_2) \\ - \end{array}\right), \quad (14)$$

where $G_{p,q}^{m,n}(\cdot)$ is the Meijer-G function, $p$ and $q$ are positive integers satisfying $\frac{p}{q} = \frac{\alpha_1}{\alpha_2}$ and $\Delta(j;x) \triangleq \frac{x}{j}, \ldots, \frac{x+j-1}{j}$. In case of the heterodyne detection, the average SNR $\mu_1$ is given by $\mu_1 = \frac{\eta \mathbb{E}[I_{2(m)}]}{\sigma_0^2}$. Regarding the IM/DD detection, the average electrical SNR $\mu_2$ is given by $\mu_2 = \frac{(\eta \mathbb{E}[I_{2(m)}])^2}{\sigma_0^2}$, while the instantaneous optical SNR is $\gamma_{2(m)} = \frac{(\eta I_{2(m)}^2)}{\sigma_0^2}$. Unifying the two detection schemes and applying the transformation of the random variable $\gamma_{2(m)} = \frac{(\eta I_{2(m)})^r}{\sigma_0^2}$, the unified PDF of the $m$-th instantaneous SNR $\gamma_{2(m)}$ can be expressed as follows

$$f_{\gamma_{2(m)}}(\gamma) = \frac{\xi^2 p^{m_2-\frac{1}{2}} q^{m_1-\frac{1}{2}} (2\pi)^{1-\frac{p+q}{2}}}{r\Gamma(m_1)\Gamma(m_2)\gamma} \times G_{p+q+\alpha_2 p, \alpha_2 p}^{0,p+q+\alpha_2 p}\left(\frac{p^p q^q \Omega_1^q \Omega_2^p}{m_1^p m_2^q}(A_0\ I_l)^{\alpha_2 p}\left(\frac{\mu_r}{\gamma}\right)^{\frac{\alpha_2 p}{r}} \middle| \begin{array}{c} \kappa_1 \\ \kappa_2 \end{array}\right), \quad (15)$$

where $\kappa_1 = \Delta(\alpha_2 p : 1-\xi^2),\ \Delta(q:1-m_1),\ \Delta(p:1-m_2)$, and $\kappa_2 = \Delta(\alpha_2 p : -\xi^2)$.

The average SNR $\overline{\gamma}_r$ can be expressed as follows:

$$\overline{\gamma}_r = \frac{\mathbb{E}[I^r]}{\mathbb{E}[I]^r}\mu_r, \quad (16)$$

The average electrical SNR $\mu_r$ can be expressed as follows:

$$\mu_r = \frac{\eta^r \mathbb{E}[I]^r}{\sigma_0^2}, \quad (17)$$

After some mathematical manipulation, the CDF can be expressed as follows:

$$F_{\gamma_{2(m)}}(\gamma) = \frac{\xi^2 p^{m_2-\frac{3}{2}} q^{m_1-\frac{1}{2}} (2\pi)^{1-\frac{p+q}{2}}}{\alpha_2 \Gamma(m_1)\Gamma(m_2)} \times G_{p+q+2\alpha_2 p, 2\alpha_2 p}^{\alpha_2 p, p+q+\alpha_2 p}\left(\frac{p^p q^q \Omega_1^q \Omega_2^p}{m_1^p m_2^q}(A_0\ I_l)^{\alpha_2 p}\left(\frac{\mu_r}{\gamma}\right)^{\frac{\alpha_2 p}{r}} \middle| \begin{array}{c} \kappa_3 \\ \kappa_4 \end{array}\right), \quad (18)$$

TABLE I
PARAMETERS OF THE FSO PART

| Parameter | Definition |
|---|---|
| $\sigma$ | Weather attenuation |
| $\sigma_s^2$ | Jitter variance |
| $\sigma_R^2$ | Rytov variance |
| $k$ | Wave number |
| $\lambda$ | Wavelength |
| $\xi$ | Pointing error coefficient |
| $\omega_0$ | Beam waist at the relay |
| $\omega_L$ | Beam waist |
| $\omega_{Leq}$ | Equivalent beam waist |
| $L$ | Length of the optical link |
| $a$ | Radius of the receiver aperture |
| $A_0$ | Fraction of the collected power at L = 0 |
| $F_0$ | Radius of curvature |
| $C_n^2$ | Refractive index of the medium |
| $R$ | Radial displacement of the beam at the receiver |

where $\kappa_3 = \kappa_1,\ [1]_{\alpha_2 p},\ \kappa_4 = [0]_{\alpha_2 p},\ \kappa_2$, and $[x]_j$ is defined as the vector of length $j$ and its components are equal to $x$.

## III. END-TO-END SINR STATISTICS

For CSI-assisted relaying, the overall SINR ($\gamma_{e2e}$) can be expressed as follows

$$\gamma_{e2e} = \frac{\gamma_{1(m)}\gamma_{2(m)}}{\gamma_{1(m)} + \gamma_{2(m)} + \gamma_{2(m)}\gamma_R + \gamma_R + 1}$$

$$= \frac{\gamma_{1(m)}^{\text{eff}} \gamma_{2(m)}}{\gamma_{1(m)}^{\text{eff}} + \gamma_{2(m)} + 1}, \quad (19)$$

where $\gamma_{1(m)}^{\text{eff}}$ is the effective RF SNR including both the interferer and the RF fadings, which can be expressed as

$$\gamma_{1(m)}^{\text{eff}} = \frac{\gamma_{1(m)}}{\gamma_R + 1}, \quad (20)$$

Given that $\gamma_{1(m)}$ and $\gamma_R$ are independent, the CDF of $\gamma_{1(m)}^{\text{eff}}$ is given by

$$F_{\gamma_{1(m)}^{\text{eff}}}(\gamma) = \Pr[\gamma_{e2e} \leq \gamma] = \Pr[\gamma_{1(m)} \leq \gamma(1+\gamma_R)]$$

$$= \int_0^\infty F_{\gamma_{1(m)}}(\gamma(1+\gamma_R)) f_{\gamma_R}(\gamma_R)\ d\gamma_R, \quad (21)$$

Using the identity [57, eq. (3.381.4)], and after some mathematical manipulations, the CDF of $\gamma_{1(m)}^{\text{eff}}$ can be expressed as follows

$$F_{\gamma_{1(m)}^{\text{eff}}}(\gamma) = \sum_{n=0}^{m-1}\sum_{i=0}^{j(m_{SR}-1)}\sum_{v=0}^{i}\mathcal{A}_0\left[1 - \sum_{l=0}^{m_{SR}+v-1}\sum_{s=0}^{l}\mathcal{A}_2\gamma^l\right.$$
$$\left. \times (\mathcal{A}_1\gamma + \beta_R)^{-(s+m_R)}e^{-\mathcal{A}_1\gamma}\right], \quad (22)$$

where $\mathcal{A}_0$, $\mathcal{A}_1$, and $\mathcal{A}_2$ are given by

$$\mathcal{A}_0 = \binom{M}{m}\binom{m-1}{n}\binom{i}{v}$$
$$\times \frac{m\,\Xi_{m_{\text{SR}}-1}^{i,j}\Gamma(m_{\text{SR}}+i)(-1)^n \rho^v (1-\rho)^{i-v}}{\Gamma(m_{\text{SR}})[1+j(1-\rho)^i](j+1)^{m_{\text{SR}}+v}}, \quad (23)$$

$$\mathcal{A}_1 = \frac{m_{\text{SR}}(j+1)}{[1+j(1-\rho)]\overline{\gamma}_1}, \quad (24)$$

$$\mathcal{A}_2 = \binom{l}{s}\frac{\beta_{\text{R}}^{m_{\text{R}}}\Gamma(m_{\text{R}}+s)\mathcal{A}_1^l}{l!\Gamma(m_{\text{R}})}, \quad (25)$$

### A. Cumulative Distribution Function

Since the CDF of $\gamma_{\text{e2e}}$ is not tractable, we refer to the following approximation

$$\gamma_{\text{e2e}} \cong \frac{\gamma_{1(\text{m})}^{\text{eff}}\gamma_{2(\text{m})}}{\gamma_{1(\text{m})}^{\text{eff}}+\gamma_{2(\text{m})}} \cong \min(\gamma_{1(\text{m})}^{\text{eff}},\gamma_{2(\text{m})}), \quad (26)$$

The approximate CDF can be expressed as follows

$$F_{\gamma_{\text{e2e}}}(\gamma) = 1 - \Pr(\min(\gamma_{1(\text{m})}^{\text{eff}},\gamma_{2(\text{m})}) \geq \gamma)$$
$$= F_{\gamma_{1(\text{m})}}^{\text{eff}}(\gamma) + F_{\gamma_{2(\text{m})}}(\gamma) - F_{\gamma_{1(\text{m})}}^{\text{eff}}(\gamma)F_{\gamma_{2(\text{m})}}(\gamma), \quad (27)$$

### B. Probability Density Function

After some mathematical manipulations, the PDF of $\gamma_{1(\text{m})}^{\text{eff}}$ can be expressed as

$$f_{\gamma_{1(\text{m})}}(\gamma) = \sum_{n=0}^{m-1}\sum_{i=0}^{j(m_{\text{SR}}-1)}\sum_{v=0}^{i}\sum_{u=0}^{m_{\text{SR}}+v}\mathcal{A}_5 \gamma^{m_{\text{SR}}+v-1}e^{-\mathcal{A}_1\gamma}, \quad (28)$$

where $\mathcal{A}_5$ is given by

$$\mathcal{A}_5 = \binom{M}{m}\binom{m-1}{n}\binom{i}{v}\binom{m_{\text{SR}}+v}{u}\left(\frac{m_{\text{SR}}}{\overline{\gamma}_1}\right)^{m_{\text{SR}}+v}$$
$$\times \frac{m(-1)^n \beta_{\text{R}}^{m_{\text{R}}}\rho^v(1-\rho)^{i-v}}{[1+j(1-\rho)]^{m_{\text{SR}}+v+i}(\mathcal{A}_1+\beta_{\text{R}})^{m_{\text{R}}+u-1}}$$
$$\times \frac{\Gamma(m_{\text{SR}}+i)\Gamma(m_{\text{R}}+u)}{\Gamma(m_{\text{SR}})\Gamma(m_{\text{R}})\Gamma(m_{\text{SR}}+v)}, \quad (29)$$

After deriving the CDF (27), the PDF can be expressed as follows

$$f_{\gamma_{\text{e2e}}}(\gamma) = f_{\gamma_{1(\text{m})}}(\gamma) + f_{\gamma_{2(\text{m})}}(\gamma) - f_{\gamma_{1(\text{m})}}(\gamma)F_{\gamma_{2(\text{m})}}(\gamma)$$
$$- F_{\gamma_{1(\text{m})}}(\gamma)f_{\gamma_{2(\text{m})}}(\gamma), \quad (30)$$

To simplify the derivation, we reformulate the CDF of $\gamma_{2(\text{m})}$ as follows

$$F_{\gamma_{2(\text{m})}}(\gamma) = \mathcal{A}_3\,G_{p+q+2\alpha_2p,2\alpha_2p}^{\alpha_2p,p+q+\alpha_2p}\left(\mathcal{A}_4\gamma^{-\frac{\alpha_2p}{r}}\,\bigg|\,\begin{matrix}\kappa_3\\\kappa_4\end{matrix}\right), \quad (31)$$

where $\mathcal{A}_3$ and $\mathcal{A}_4$ are defined by

$$\mathcal{A}_3 = \frac{\xi^2 p^{m_2-\frac{3}{2}}q^{m_1-\frac{1}{2}}(2\pi)^{1-\frac{p+q}{2}}}{\alpha_2 \Gamma(m_1)\Gamma(m_2)}, \quad (32)$$

$$\mathcal{A}_4 = \left(\frac{q\Omega_1}{m_1}\right)^q\left(\frac{p\Omega_2}{m_2}\right)^p (A_0 I_l)^{\alpha_2p}\mu_r^{\frac{\alpha_2p}{r}}, \quad (33)$$

We also reformulate the PDF of $\gamma_{2(\text{m})}$ as follows

$$f_{\gamma_{2(\text{m})}}(\gamma) = \frac{\mathcal{A}_6}{\gamma}G_{p+q+\alpha_2p,\alpha_2p}^{0,p+q+\alpha_2p}\left(\mathcal{A}_4\gamma^{-\frac{\alpha_2p}{r}}\,\bigg|\,\begin{matrix}\kappa_1\\\kappa_2\end{matrix}\right), \quad (34)$$

where $\mathcal{A}_6$ is given by

$$\mathcal{A}_6 = \frac{\xi^2 p^{m_2-\frac{1}{2}}q^{m_1-\frac{1}{2}}(2\pi)^{1-\frac{p+q}{2}}}{r\Gamma(m_1)\Gamma(m_2)}, \quad (35)$$

### C. Moments

The $\nu$-th moment is defined as follows:

$$\mathbb{E}[\gamma^\nu] = \int_0^\infty \gamma^\nu f_\gamma(\gamma)d\gamma, \quad (36)$$

After replacing the PDF expression (30) in Eq. (36), the moments are expressed as follows

$$\mathbb{E}[\gamma^\nu] = \mathcal{I}_1 + \mathcal{I}_2 - \mathcal{I}_3 - \mathcal{I}_4, \quad (37)$$

Using the identity [57, eq. (3.381.4)], the term $\mathcal{I}_1$ can be expressed as follows

$$\mathcal{I}_1 = \int_0^\infty \gamma^\nu f_{\gamma_{1(\text{m})}}(\gamma)d\gamma$$
$$= \sum_{n=0}^{m-1}\sum_{i=0}^{j(m_{\text{SR}}-1)}\sum_{v=0}^{i}\sum_{u=0}^{m_{\text{SR}}+v}\frac{\mathcal{A}_5}{\mathcal{A}_1^{m_{\text{SR}}+\nu+v}}\Gamma(m_{\text{SR}}+v+\nu), \quad (38)$$

After changing the variable of integration ($x = \gamma^{-\frac{\alpha_2p}{r}}$) and using the identity [58, eq. (2.24.2.1)], the term $\mathcal{I}_2$ can be obtained by eq. (39), as shown at the bottom of this page.

After changing the variable of integration ($x = \gamma^{-1}$) and using the identity [58, eq. (2.24.1.1)], the term $\mathcal{I}_3$ is given by Eq. (40), as shown at the top of the next page, where $\zeta_1 = \sum_{j=1}^{2\alpha_2p}\kappa_{4,j} - \sum_{j=1}^{p+q+2\alpha_2p}\kappa_{3,j} + \frac{p+q}{2} + 1,$ $\kappa_5 = \Delta(r:\alpha_2p:1-\xi^2),$ $\Delta(r:q:1-m_1),$ $\Delta(r:p:1-m_2),$ $\Delta(r:\alpha_2p:1),$ and $\kappa_6 = \Delta(\alpha_2p:m_{\text{SR}}+v+\nu),$ $\Delta(r:\alpha_2p:0), \Delta(r:\alpha_2p:-\xi^2).$

$$\mathcal{I}_2 = \int_0^\infty \gamma^\nu f_{\gamma_{2(\text{m})}}(\gamma)d\gamma = \frac{r\mathcal{A}_6}{\alpha_2 p\mathcal{A}_4^{\frac{r\nu}{\alpha_2p}+2}}\frac{\prod_{j=1}^{p+q+\alpha_2p}\Gamma\left(-\frac{r\nu}{\alpha_2p}-\kappa_{1,j}-1\right)}{\prod_{j=1}^{\alpha_2p}\Gamma\left(-\frac{r\nu}{\alpha_2p}-\kappa_{2,j}-1\right)}, \quad (39)$$

$$\mathcal{I}_3 = \int_0^\infty \gamma^\nu f_{\gamma_{1(m)}}(\gamma) F_{\gamma_{2(m)}}(\gamma) d\gamma = \sum_{n=0}^{m-1} \sum_{i=0}^{j(m_{SR}-1)} \sum_{v=0}^{i} \sum_{u=0}^{m_{SR}+v} \frac{\mathcal{A}_3 \mathcal{A}_5 r^{\zeta_1} (\alpha_2 p)^{\frac{2(m_{SR}+v+\nu)-1}{2}}}{\mathcal{A}_1^{m_{SR}+v+\nu} (2\pi)^{\frac{\alpha_2 p+(r-1)(p+q)-1}{2}}}$$

$$\times G_{r(p+q+\alpha_2 p),(r+1)\alpha_2 p}^{\alpha_2 p, r(p+q+\alpha_2 p)} \left( (\mathcal{A}_4 r^{p+q})^r \left(\frac{\mathcal{A}_1}{\alpha_2 p}\right)^{\alpha_2 p} \bigg| \begin{matrix} \kappa_5 \\ \kappa_6 \end{matrix} \right), \tag{40}$$

$$\mathcal{I}_4 = \int_0^\infty \gamma^\nu F_{\gamma_{1(m)}}(\gamma) f_{\gamma_{2(m)}}(\gamma) d\gamma = \sum_{n=0}^{m-1} \sum_{i=0}^{j(m_{SR}-1)} \sum_{v=0}^{i} \mathcal{A}_0 \left[ \mathcal{I}_2 + \frac{r\mathcal{A}_6}{\alpha_2 p} \sum_{l=0}^{m_{SR}+v-1} \sum_{s=0}^{l} \frac{\mathcal{A}_2 \mathcal{A}_1^{-(n+l)}}{\beta_R^{s+m_R} \Gamma(s+m_R)} \right.$$

$$\left. \times H_{1,0:1,1:p+q+\alpha_2 p,\alpha_2 p}^{0,1:1,1:1,0,p+q+\alpha_2 p} \left( \begin{matrix} (1-l-\nu;1,1) \\ - \end{matrix} \bigg| \begin{matrix} (1-s-m_R,1) \\ (0,1) \end{matrix} \bigg| \begin{matrix} (\kappa_1,[-\frac{r}{\alpha_2 p}]_{p+q+\alpha_2 p}) \\ (\kappa_2,[-\frac{r}{\alpha_2 p}]_{\alpha_2 p}) \end{matrix} \bigg| \begin{matrix} \frac{1}{\beta_R}, \frac{\mathcal{A}_4^{-\frac{r}{\alpha_2 p}}}{\mathcal{A}_1} \end{matrix} \right) \right], \tag{41}$$

$$\mathcal{J}_1 = \sum_{n=0}^{m-1} \sum_{i=0}^{j(m_{SR}-1)} \sum_{v=0}^{i} \mathcal{A}_0 \left[ 1 - \sum_{l=0}^{m_{SR}+v-1} \sum_{s=0}^{l} t \mathcal{A}_2 \beta_R^{l-m_R-s+1} \frac{\Gamma(l+1)}{\mathcal{A}_1^{l+1}} \Psi\left(l+1, l+2-m_R-s; \frac{\beta_R(\mathcal{A}_1+t)}{\mathcal{A}_1}\right) \right], \tag{43}$$

$$\mathcal{J}_2 = \frac{\mathcal{A}_3 r^{\zeta_1} \sqrt{\alpha_2 p}}{(2\pi)^{\frac{\alpha_2 p+(r-1)(p+q)-1}{2}}} G_{r(p+q+\alpha_2 p),(r+1)\alpha_2 p}^{\alpha_2 p, r(p+q+\alpha_2 p)} \left( (\mathcal{A}_4 r^{p+q})^r \left(\frac{t}{\alpha_2 p}\right)^{\alpha_2 p} \bigg| \begin{matrix} \kappa_5 \\ \kappa_7 \end{matrix} \right), \tag{44}$$

$$\mathcal{J}_3 = \sum_{n=0}^{m-1} \sum_{i=0}^{j(m_{SR}-1)} \sum_{v=0}^{i} t \mathcal{A}_0 \left[ \mathcal{J}_2 + \sum_{l=0}^{m_{SR}+v-1} \sum_{s=0}^{l} \frac{r \mathcal{A}_2 \mathcal{A}_3}{\alpha_2 p \beta_R^{s+m_R} (\mathcal{A}_1+t)^{l+1} \Gamma(m_R+s)} \right.$$

$$\left. \times H_{1,0:1,1:p+q+\alpha_2 p,\alpha_2 p}^{0,1:1,1:1,0,p+q+\alpha_2 p} \left( \begin{matrix} (-l;1,1) \\ - \end{matrix} \bigg| \begin{matrix} (1-s-m_R,1) \\ (0,1) \end{matrix} \bigg| \begin{matrix} (\kappa_1,[-\frac{r}{\alpha_2 p}]_{p+q+\alpha_2 p}) \\ (\kappa_2,[-\frac{r}{\alpha_2 p}]_{\alpha_2 p}) \end{matrix} \bigg| \begin{matrix} \frac{\mathcal{A}_1}{\beta_R(\mathcal{A}_1+t)}, \frac{\mathcal{A}_4^{-\frac{r}{\alpha_2 p}}}{\mathcal{A}_1+t} \end{matrix} \right) \right], \tag{45}$$

$$G_{p+q+2\alpha_2 p, 2\alpha_2 p}^{\alpha_2 p, p+q+\alpha_2 p} \left( \mathcal{A}_4 \gamma^{-\frac{\alpha_2 p}{r}} \bigg| \begin{matrix} \kappa_3 \\ \kappa_4 \end{matrix} \right) \underset{\mu_r \gg 1}{\cong} \sum_{k=1}^{p+q+\alpha_2 p} \frac{\prod_{j=1, j \neq k}^{p+q+\alpha_2 p} \Gamma(\kappa_{3,k}-\kappa_{3,j}) \prod_{j=1}^{\alpha_2 p} \Gamma(1-\kappa_{3,k}+\kappa_{4,j}) \left(\mathcal{A}_4 \gamma^{-\frac{\alpha_2 p}{r}}\right)^{\kappa_{3,k}-1}}{\prod_{j=1\alpha_2 p+1}^{2\alpha_2 p} \Gamma(\kappa_{3,k}-\kappa_{4,j}) \prod_{j=p+q+\alpha_2 p+1}^{p+q+2\alpha_2 p} \Gamma(\kappa_{3,j}-\kappa_{3,k}+1)}, \tag{47}$$

---

After reproducing the same derivation steps as $\mathcal{I}_2$ and using the identities [58, eq. (8.4.3.1)], [59, eq. (07.35.03.0001.01), (07.35.26.0003.01)], and [60, eq. (2.3)], the term $\mathcal{I}_4$ can be expressed by Eq. (41), as shown at the top of this page, where $H_{p_1,q_1:p_2,q_2:p_3,q_3}^{m_1,n_1:m_2,n_2:m_3,n_3}(\cdot)$ is the bivariate Fox-H function. An efficient implementation of this function is provided by [54] and [55].

### D. Moment Generating Function

The moment generating function can be expressed in terms of the CDF as follows [44, eq. (12)]:

$$\mathcal{M}_\gamma(t) = \mathbb{E}\left[e^{t\gamma}\right] = t\int_0^\infty e^{t\gamma} F_\gamma(\gamma) d\gamma, \tag{42}$$

After replacing the CDF (27) in Eq. (42), the MGF can be expressed as the summation of three terms $\mathcal{J}_1$, $\mathcal{J}_2$, and $\mathcal{J}_3$.

Using the identities [57, eq. (3.381.4)], [63, eq. (2.3.6.9)] and after some mathematical manipulation, the term $\mathcal{J}_1$ can be given by Eq. (43), as shown at the top of this page, where $\Psi(\cdot : \cdot ; \cdot)$ is the Tricomi confluent hypergeometric function.

After applying the identity [58, eq. (2.24.3.1)], the term $\mathcal{J}_2$ can be obtained by Eq. (44), as shown at the top of this page, where $\kappa_7 = \Delta(\alpha_2 p : 1)$, $\Delta(r : \alpha_2 p : 0)$, $\Delta(r : \alpha_2 p : -\xi^2)$.

Following the same derivation steps for $\mathcal{I}_4$, the term $\mathcal{J}_3$ can be given by Eq. (45), as shown at the top of this page.

## IV. PERFORMANCE ANALYSIS

### A. End-to-End Outage Probability

The end-to-end outage probability is the probability that the overall SINR falls below a given threshold $\gamma_T$. For CSI-assisted relaying, the outage probability can be given using (27).

$$P_{out}(\gamma_T) = \Pr[\gamma \leq \gamma_T] = F_{\gamma_{e2e}}(\gamma_T), \tag{46}$$

### B. High SNR Analysis

To get the diversity gain $G_d$, we derive the asymptotic high SNR by expanding the Meijer-G function in (31) using [59, eq. (07.34.06.0044.01)]. The expression is given by Eq. (46). For infinite RF, and FSO average SNR, and after applying partial fraction expansion on (22), it can be shown that the diversity gain $G_d$ is given by Eq. (48).

Note that the diversity gain for partial relay selection of the RF branches is equal to $m_{SR}$ regardless of the correlation coefficient value $\rho$, unlike the case of opportunistic relay selection protocol wherein the correlation affects the diversity

TABLE II
PARAMETERS OF BINARY MODULATIONS

| Modulation | $\delta$ | $\tau$ |
|---|---|---|
| Coherent Binary Frequency Shift Keying (CBFSK) | 0.5 | 0.5 |
| Non-Coherent Binary Frequency Shift Keying (NBFSK) | 0.5 | 1 |
| Coherent Binary Phase Shift Keying (CBPSK) | 1 | 0.5 |
| Differential Binary Phase Shift Keying (DBPSK) | 1 | 1 |

gain.

$$G_{\mathrm{d}} = \min\left(\frac{\xi^2}{r}, \frac{m_1\alpha_1}{r}, \frac{m_2\alpha_2}{r}, m_{\mathrm{SR}}\right), \quad (48)$$

### C. Higher-Order Amount of Fading

The amount of fading is mathematically defines as follows

$$AF_\gamma^{(\nu)} = \frac{\mathbb{E}[\gamma^\nu]}{\mathbb{E}[\gamma]^\nu} - 1, \quad (49)$$

Replacing (37) in (49) yields to the $\nu$-th order of the amount of fading.

### D. Average Bit Error Probability

For the most binary modulations, the bit error probability is expressed as follows

$$\overline{P_{\mathrm{e}}} = \frac{\delta^\tau}{2\Gamma(\tau)} \int_0^\infty \gamma^{\tau-1} e^{-\delta\gamma} F_\gamma(\gamma) d\gamma, \quad (50)$$

where $\tau$ and $\delta$ are the parameters of the modulation, which can be summarized in Table II.

For CSI-assisted relaying of the proposed system, the bit error rate can be given by replacing (27) in (50). In this case, it can be expressed as follows

$$\overline{P_{\mathrm{e}}} = \mathcal{T}_1 + \mathcal{T}_2 - \mathcal{T}_3, \quad (51)$$

Using the identities [57, eq. (3.381.4)] and [63, eq. (2.3.6.9)], the term $\mathcal{T}_1$ can be derived as follows

$$\mathcal{T}_1 = \frac{\delta^\tau}{2\Gamma(\tau)} \int_0^\infty \gamma^{\tau-1} e^{-\delta\gamma} F_{\gamma_{1(\mathrm{m})}}(\gamma) d\gamma = \sum_{n=0}^{m-1} \sum_{i=0}^{j(m_{\mathrm{SR}}-1)} \sum_{v=0}^{i} \frac{\mathcal{A}_0}{2t}$$
$$\times \left[1 - \sum_{l=0}^{m_{\mathrm{SR}}+v-1} \sum_{s=0}^{l} \frac{\mathcal{A}_2 \delta^\tau \beta_{\mathrm{R}}^{l+\tau-m_{\mathrm{R}}-s} \Gamma(l+1)}{\mathcal{A}_1^{l+\tau}\Gamma(\tau)} \right.$$
$$\left. \times \Psi\left(l+\tau, l+\tau-m_{\mathrm{R}}-s+1; \frac{\beta_{\mathrm{R}}(\mathcal{A}_1+\delta)}{\mathcal{A}_1}\right)\right], \quad (52)$$

After changing the variable of integration ($x = \gamma^{-1}$), and using the identities [59, eqs. (01.03.26.0004.01), (07.34.16.0002.01)], and [58, eq. (2.24.1.1)], the term $\mathcal{T}_2$ can derived as follows

$$\mathcal{T}_2 = \frac{\delta^\tau}{2\Gamma(\tau)} \int_0^\infty \gamma^{\tau-1} e^{-\delta\gamma} F_{\gamma_{2(\mathrm{m})}}(\gamma) d\gamma$$
$$= \frac{\mathcal{A}_3 r^{\zeta_1}(\alpha_2 p)^{(l+\tau-1)}}{2(2\pi)^{\frac{\alpha_2 p+(r-1)(p+q)-1}{2}}\Gamma(\tau)}$$

$$\times G_{r(p+q+\alpha_2 p),(r+1)\alpha_2 p}^{\alpha_2 p, r(p+q+\alpha_2 p)} \left( (\mathcal{A}_4 r^{p+q})^r \left(\frac{\delta}{\alpha_2 p}\right)^{\alpha_2 p} \middle| \begin{matrix} \kappa_5 \\ \kappa_8 \end{matrix} \right), \quad (53)$$

where $\kappa_8 = \Delta(\alpha_2 p : \tau)$, $\Delta(r : \alpha_2 p : 0)$, $\Delta(r : \alpha_2 p : -\xi^2)$.

After reproducing the same derivation steps for $\mathcal{I}_4$, term $\mathcal{T}_3$ can be expressed by (57), as shown at the bottom of the next page.

### E. Ergodic Capacity

The channel capacity, expressed in (bit/s/Hz), is defined as the maximum error-free data rate transmitted by the system. It can be written as follows

$$\overline{C} = \mathbb{E}[\log_2(1+\varpi\gamma)] = \int_0^\infty \log_2(1+\varpi\gamma) f_\gamma(\gamma) d\gamma, \quad (54)$$

$\varpi = 1$ or $\frac{e}{2\pi}$, respectively, for heterodyne and IM/DD detection.

After replacing the PDF (30) in (54), the ergodic capacity can be expressed as follows

$$\overline{C} = \mathcal{C}_1 + \mathcal{C}_2 - \mathcal{C}_3 - \mathcal{C}_4, \quad (55)$$

After applying [58, eqs. (8.4.6.5), (2.24.3.1)], and [59, eq. (07.35.03.0001.01), (07.35.26.0003.01)], the term $\mathcal{C}_1$ can be derived in terms of the univariate Fox-H function (58), as shown at the bottom of the next page.

Reproducing the same procedures for $\mathcal{C}_1$, the term $\mathcal{C}_2$ can be obtained by (59), as shown at the bottom of the next page.

Using the identities [58, eqs. (8.4.3.1), (8.4.6.5)], [59, eq. (07.35.03.0001.01), (07.35.26.0003.01)], and [60, eq. (2.3)] and after some mathematical manipulations, the term $\mathcal{C}_3$ can be derived by (60), as shown at the bottom of the next page.

The term $\mathcal{C}_4$ can be derived in terms of the trivariate Fox-H function (61), as shown at the bottom of the next page,

*Proof:* The derivation steps of $\mathcal{C}_4$ are provided in Appendix A. The Python implementation of the multivariate Fox-$H$ function is given by [64]. ∎

### F. End-to-End Outage Capacity

The outage capacity is defined as the probability that the overall throughput falls below a given outage rate $\mathcal{C}_{\mathrm{T}}$. This metric is very important since it describes clearly the average throughput outage of the proposed system. Mathematically, the outage rate can be expressed as follows:

$$\mathcal{R}(\mathcal{C}_{\mathrm{T}}) = \Pr[\overline{C} \leq \mathcal{C}_{\mathrm{T}}] = F_\gamma\left(\frac{2^{\mathcal{C}_{\mathrm{T}}}-1}{\varpi}\right) \quad (56)$$

After replacing the CDF (27) in (56), the outage rate is finally derived. Note that $\overline{\kappa} = [1]_{\mathrm{length}(\kappa)} - \kappa$.

## V. NUMERICAL RESULTS AND DISCUSSIONS

In this section, we compare the analytical expressions of the system performance against the Monte Carlo simulations. The correlated RF CSI is generated using relation (1), while the atmospheric turbulences samples are generated using the product of two independent random variables ($I_a = I_{aX} \times I_{aY}$) following the Generalized Gamma distribution. In addition, the pointing error samples are generated by firstly generating the radial displacement $R$ following the Rayleigh distribution with scale equal to the jitter standard deviation ($\sigma_s$) and then we generate the samples using (13). Since the path loss is deterministic, it can be generated using relation (12). Table III summarizes the main simulation parameters.

TABLE III
MAIN SIMULATION PARAMETERS

| Parameter | Value |
| --- | --- |
| $L$ | 1 km |
| $\lambda$ | 1550 nm |
| $F_0$ | -10 m |
| $a$ | 5 cm |
| $\omega_0$ | 5 mm |
| $\sigma_s$ | 3.75 cm |
| $\sigma$ | 0.5 dB/km |
| $p$ | 2 |
| $q$ | 2 |
| $m_1$ | 3 |
| $m_2$ | 3 |
| $M_R$ | 5 |
| $m_R$ | 5 |

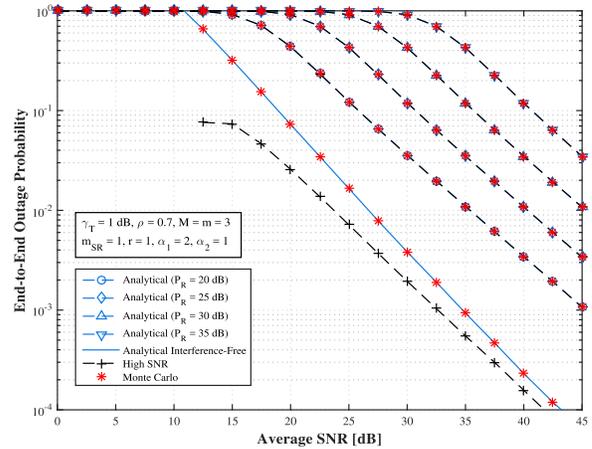

Fig. 2. Effects of the interferers' powers on the outage probability.

Fig. 2 shows the end-to-end outage performance for various profiles of interferers. As a special case, we assume that the RF channels experience Rayleigh fading ($m_{SR} = 1$). We observe that more interferers's powers yields worse outage performance. In this case, to improve the network coverage and scalability in farthest areas, it is better to implement useful techniques to eliminate or reduce the interference impacts such as the partial interference cancellation. Interference-free case illustrates the best performance compared to the other ones.

The impacts of the pointing error on the outage probability for heterodyne and IM/DD detection modes are illustrated by Fig. 3. As expected, the system works better under the

$$\mathcal{T}_3 = \frac{\delta^\tau}{2\Gamma(\tau)} \int_0^\infty \gamma^{\tau-1} e^{-\delta\gamma} F_{\gamma_{1(m)}}(\gamma) F_{\gamma_{2(m)}}(\gamma) d\gamma = \sum_{n=0}^{m-1} \sum_{i=0}^{j(m_{SR}-1)} \sum_{v=0}^{i} \mathcal{A}_0 \left[ \mathcal{T}_2 + \sum_{l=0}^{m_{SR}+v-1} \sum_{s=0}^{l} \frac{r\delta^\tau}{2\alpha_2 p \beta_R^{m_R+s}(\mathcal{A}_1+\delta)^{l+\tau}} \right.$$
$$\left. \times \frac{\mathcal{A}_2 \mathcal{A}_3}{\Gamma(\tau)\Gamma(m_R+s)} H_{1,0:1,1:p+q+2\alpha_2 p,2\alpha_2 p}^{0,1:1,1:\alpha_2 p,p+q+\alpha_2 p} \left( \begin{matrix} (1-l-\tau;1,1) \\ - \end{matrix} \middle| \begin{matrix} (1-s-m_R,1) \\ (0,1) \end{matrix} \middle| \begin{matrix} (\kappa_3,[-\frac{r}{\alpha_2 p}]_{p+q+2\alpha_2 p}) \\ (\kappa_4,[-\frac{r}{\alpha_2 p}]_{2\alpha_2 p}) \end{matrix} \middle| \frac{\mathcal{A}_1}{\beta_R(\mathcal{A}_1+\delta)}, \frac{\mathcal{A}_4^{-\frac{r}{\alpha_2 p}}}{\mathcal{A}_1+\delta} \right) \right], \tag{57}$$

$$\mathcal{C}_1 = \int_0^\infty \log_2(1+\varpi\gamma) f_{\gamma_{1(m)}}(\gamma) d\gamma = \sum_{n=0}^{m-1} \sum_{i=0}^{j(m_{SR}-1)} \sum_{v=0}^{i} \sum_{u=0}^{m_{SR}+v} \frac{\mathcal{A}_5}{\log(2)\mathcal{A}_1^{m_{SR}+v}} H_{3,2}^{1,3} \left( \begin{matrix} (1-m_{SR}-v,1)(1,1)(1,1) \\ (1,1)(0,1) \end{matrix} \middle| \frac{\varpi}{\mathcal{A}_1} \right), \tag{58}$$

$$\mathcal{C}_2 = \int_0^\infty \log_2(1+\varpi\gamma) f_{\gamma_{2(m)}}(\gamma) d\gamma = -\frac{r\mathcal{A}_6}{\alpha_2 p \log(2)} H_{p+q+\alpha_2 p+2,2\alpha_2 p+2}^{2,p+q+\alpha_2 p+1} \left( \begin{matrix} (\kappa_1,[-\frac{r}{\alpha_2 p}]_{p+q+\alpha_2 p+2})(0,1)(0,1) \\ (0,1)(0,1)(\kappa_2,[-\frac{r}{\alpha_2 p}]_{\alpha_2 p+2}) \end{matrix} \middle| \frac{\mathcal{A}_4^{-\frac{r}{\alpha_2 p}}}{\varpi} \right), \tag{59}$$

$$\mathcal{C}_3 = \int_0^\infty \log_2(1+\varpi\gamma) f_{\gamma_{1(m)}}(\gamma) F_{\gamma_{2(m)}}(\gamma) d\gamma = -\frac{r}{\alpha_2 p \log(2)} \sum_{n=0}^{m-1} \sum_{i=0}^{j(m_{SR}-1)} \sum_{v=0}^{i} \sum_{u=0}^{m_{SR}+v} \frac{\mathcal{A}_3 \mathcal{A}_5}{\mathcal{A}_1^{m_{SR}+v}}$$
$$\times H_{1,0:2,2:p+q+2\alpha_2 p,2\alpha_2 p}^{0,1:1,2:\alpha_2 p,p+q+\alpha_2 p} \left( \begin{matrix} (1-m_{SR}-v;1,1) \\ - \end{matrix} \middle| \begin{matrix} (1,1)(1,1) \\ (1,1)(0,1) \end{matrix} \middle| \begin{matrix} (\kappa_3,[-\frac{r}{\alpha_2 p}]_{p+q+2\alpha_2 p}) \\ (\kappa_4,[-\frac{r}{\alpha_2 p}]_{2\alpha_2 p}) \end{matrix} \middle| \frac{\varpi}{\mathcal{A}_1}, \frac{1}{\mathcal{A}_1 \mathcal{A}_4^{\frac{\alpha_2 p}{r}}} \right), \tag{60}$$

$$\mathcal{C}_4 = \int_0^\infty \log_2(1+\varpi\gamma) F_{\gamma_{1(m)}}(\gamma) f_{\gamma_{2(m)}}(\gamma) d\gamma = \sum_{n=0}^{m-1} \sum_{i=0}^{j(m_{SR}-1)} \sum_{v=0}^{i} \left[ \mathcal{C}_2 - \sum_{l=0}^{m_{SR}+v-1} \sum_{s=0}^{l} \frac{\mathcal{A}_1^{l+1} \mathcal{A}_2 \mathcal{A}_3}{\beta_R^{m_R+s} \Gamma(m_R+s) \log(2)} \right.$$
$$\left. \times H_{1,0:0,2:2,2:p+q+2\alpha_2 p,2\alpha_2 p}^{0,1:2,0:2,1:0,p+q+\alpha_2 p} \left( \begin{matrix} (-l;-1,-1,-\frac{\alpha_2 p}{r}) \\ - \end{matrix} \middle| \begin{matrix} - \\ (1,-1)(0,1)(0,1) \end{matrix} \middle| \begin{matrix} (0,1)(1,1) \\ (0,1)(0,1) \end{matrix} \middle| \begin{matrix} (\overline{\kappa}_2,[-\frac{r}{\alpha_2 p}]_{\alpha_2 p}) \\ (\overline{\kappa}_1,[-\frac{r}{\alpha_2 p}]_{p+q+\alpha_2 p}) \end{matrix} \middle| \frac{\beta_R}{\mathcal{A}_1^2}, \frac{\mathcal{A}_1^{-1}}{\varpi}, \frac{\mathcal{A}_4^{-\frac{\alpha_2 p}{r}}}{\mathcal{A}_1} \right) \right], \tag{61}$$

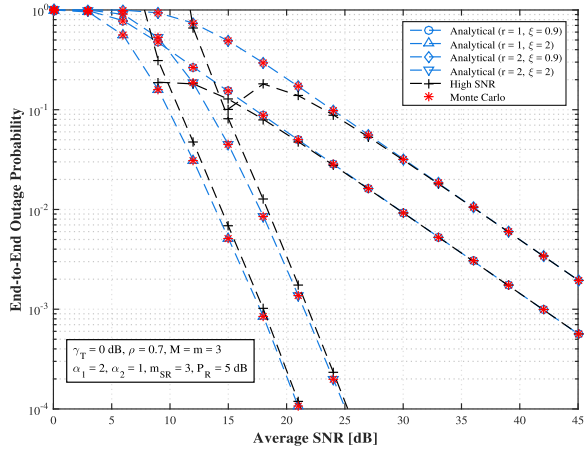

Fig. 3. Effects of the pointing error on the outage performance.

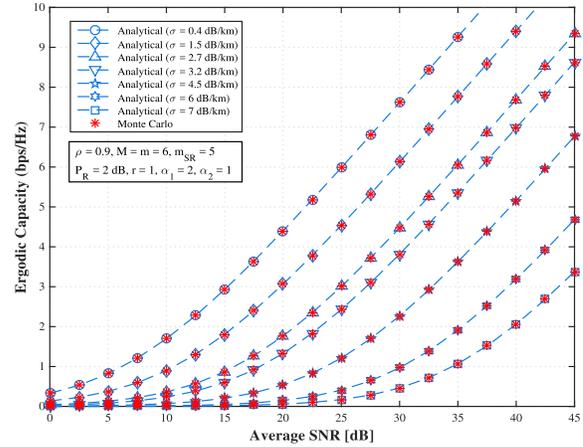

Fig. 5. Effects of the atmospheric path loss on the average capacity.

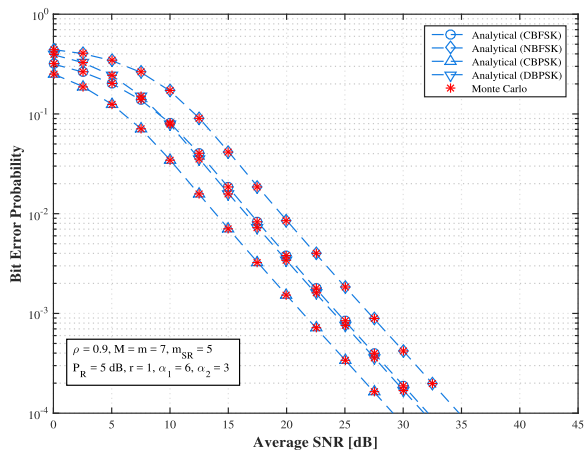

Fig. 4. Bit error probability for various binary modulation schemes.

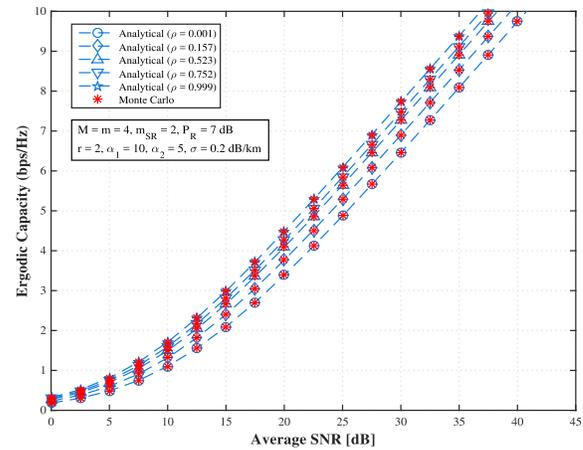

Fig. 6. Effects of the time correlation on the ergodic capacity.

coherent detection rather than IM/DD for negligeable pointing error. In addition, the system performance is very sensitive to the pointing error coefficients. In fact, we observe that as the pointing error coefficient $\xi$ decreases (severe pointing error fading), the effect becomes more pronounced and the performance gets worse. We also note that for severe pointing fading, the system performance assuming coherent detection is worse compared to the case of IM/DD for less pointing error fading. Thereby, the coverage reliability depends to a large extent on the misalignment between the relays and the front photodetector.

Fig. 4 shows the variations of the bit error probability for various binary modulation schemes. The graph shows clearly the agreement between the derived analytical results and the Monte Carlo simulation. Therefore, these results confirm the accuracy of the performance metrics derived of the proposed system. Furthermore, we note that the best performance is achieved by CBPSK, however, it becomes completely bad for NBFSK modulation.

Fig. 5 illustrates the variations of the average capacity for different values of the atmospheric weather attenuation (path loss). We observe that for lower path loss value roughly 0.4 dB/km, which describes a clear air weather, the system achieves better throughput. As the path loss becomes moderate for rainy weather around 2.7 and 4.5 dB/km, the system still operates in acceptable conditions but with lower throughput compared to the case of clear air condition. However, as the atmospheric attenuation becomes more severe, which is the case of foggy weather, the average capacity substantially gets worse. We also observe that for an average SNR around 45 dB, the throughput is roughly 3.2 bps/Hz for severe path loss ($\sigma = 7$ dB/km) while for moderate path loss ($\sigma = 4.5$ dB/km), the achievable rate is around 6.8 bps/Hz. Consequently, the effect of the atmospheric attenuation on the system throughput is substantially pronounced mainly at high SNR.

Fig. 6 illustrates the effects of the time correlation coefficient $\rho$ on the ergodic capacity. We observe that the average rate gets better as the correlation between the RF CSIs increases. In fact, for higher correlation, the source gets better estimation of the channels' coefficients and based on that, the best branch will be selected for the transmission. However, as the RF CSIs become completely outdated ($\rho \cong 0.001$), the source gets bad estimation of the channels' coefficients and hence the selection of the best branch is uncertainly achieved. Generally speaking, the throughput basically improves as the instantaneous CSI of the branch becomes stronger. Consequently, to realize a stable and satisfied average

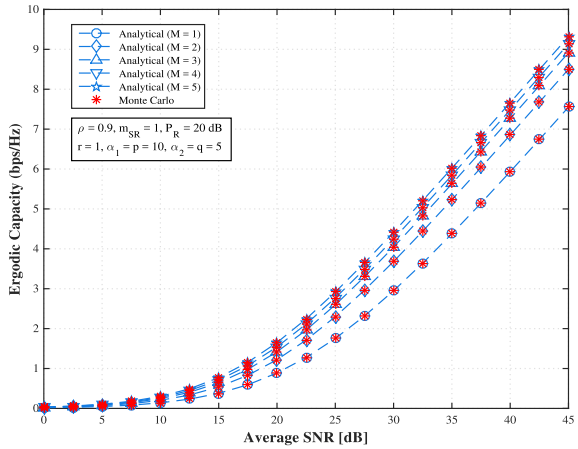

Fig. 7. Average capacity performance for various number of relays.

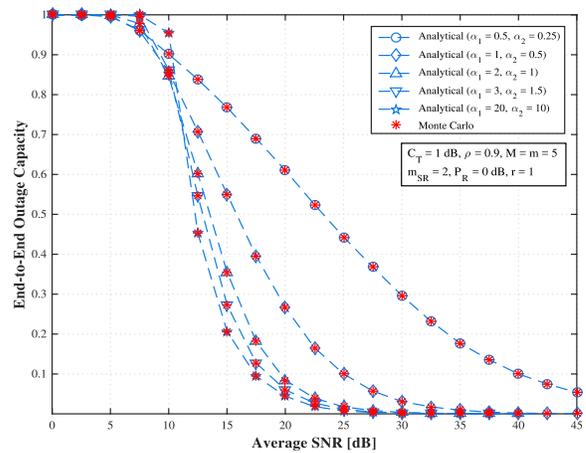

Fig. 9. End-to-end outage capacity under various turbulence conditions.

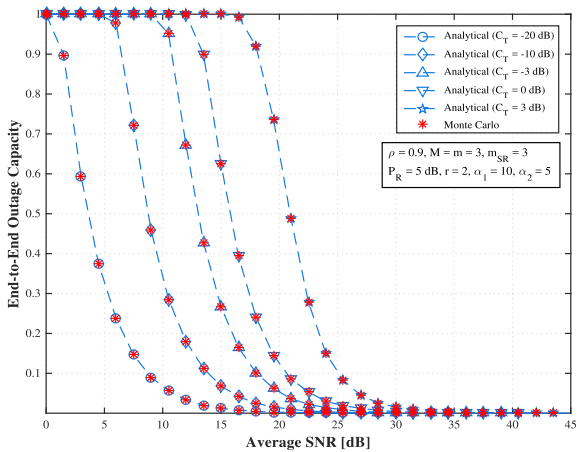

Fig. 8. Effects of the capacity threshold on the end-to-end outage rate.

throughput, a better channel estimation (higher correlation) must be achieved first.

The variations of the ergodic capacity for various number of relays is illustrated by Fig. 7. We clearly note that a large number of relays yields better throughput. In fact, increasing the number of the relays means that the source has better chance to select a branch with a stronger CSI. Therefore, to serve the densified cells without throughput perturbation, the number of the relays implemented must be large enough to deal mainly with the power shortage/outage that may occur for farther communications. Moreover, for a given throughput equal to 3 bps/Hz, the proposed system requires 30 dB and 25 dB for M = 1 and 5, respectively. Thereby, the system achieves a power gain of 5 dB.

Another important metric used to evaluate the system performance is the outage capacity whose variations with respect to the outage threshold, and the atmospheric turbulence are shown in figures 8, and 9, respectively. We clearly note in Fig. 8 that lower threshold yields better throughput coverage. However, as the throushold becomes stronger, the throughput quickly saturates and reaches the bottleneck. In addition, the variations of the outage capacity for various atmospheric turbulence conditions are illustrated by Fig. 9. As expected, weaker atmospheric turbulence conditions

(higher values of $\alpha_1$ and $\alpha_2$) yields lower outage throughput. For moderate turbulence ($\alpha_1 = 3$ and $\alpha_2 = 1.5$), the system still achieves acceptable performance but a little worse compared to the case of weak turbulence. However, as the turbulences become severe ($\alpha_1 = 0.5$ and $\alpha_2 = 0.25$), the average throughput is not stable anymore and experiences a substantial outage/shortage yielding to the worst performance. As a result, the system performance depends to a large extent on the state of the optical channel.

## VI. CONCLUSION

In this work, we proposed a dual-hop mixed RF/FSO system with multiple relays under the effects of the co-channel interference. Partial relay selection with outdated CSI is assumed as a protocol to select the best branch/relay. The results show that the coverage reliability is very sensitive to the interferers' powers. Moreover, a large number of relays and higher correlation substantially improves the system throughput under weak and moderate atmospheric conditions. However, the contributions of these parameters become less pronounced as the state of the optical channel (path loss, pointing error and atmospheric turbulences) becomes instable. The most challenging part of FSO system is the strong dependence on the severity of fading and possible enhancements are very limited. As an extention of this work, we intend to propose sophisticated techniques to mitigate or compensate for the loss introduced by the FSO channel disturbance.

## APPENDIX A
### DERIVATION OF TERM $\mathcal{C}_4$

Term $\mathcal{C}_4$ consists of two integral expressions where the left hand side is term $\mathcal{C}_2$ and the right hand side is $\mathcal{C}_5$ given by

$$\mathcal{C}_5 = \int_0^\infty \gamma^l (\mathcal{A}_1 \gamma + \beta_R)^{-(s+m_R)} \log(1 + \varpi \gamma) e^{-\mathcal{A}_1 \gamma}$$

$$\times G_{p+q+\alpha_2 p, \alpha_2 p}^{0, p+q+\alpha_2 p} \left( \mathcal{A}_4 \gamma^{-\frac{\alpha_2 p}{r}} \middle| \begin{array}{c} \kappa_1 \\ \kappa_2 \end{array} \right) d\gamma, \quad (62)$$

The argument of the Meijer-G function in the aforementioned integral must be inverted using [59, eq. (8.2.2.14)].

$$\mathcal{C}_5 = \frac{1}{\Gamma(s+m_R)\beta^{s+m_R}(2\pi i)^3} \int_{\mathcal{L}_1}\int_{\mathcal{L}_2}\int_{\mathcal{L}_3} \frac{\Gamma(-s_1)\Gamma(s+m_R+s_1)\prod_{j=1}^{p+q+\alpha_2 p}\Gamma(1-\kappa_{2,j}-s_2)\Gamma(1-s_3)\Gamma^2(s_3)}{\prod_{j=1}^{\alpha_2 p}\Gamma(1-\kappa_{1,j}-s_2)\Gamma(s_3+1)}$$
$$\times \left(\frac{\mathcal{A}_1}{\beta_R}\right)^{s_1} \mathcal{A}_4^{-s_2} \underbrace{\int_0^\infty \gamma^{l+s_1+s_3+\frac{\alpha_2 p s_2}{r}} e^{-\mathcal{A}_1 \gamma} d\gamma}_{\mathcal{I}} \; ds_1 \; ds_2 \; ds_3, \quad (70)$$

After referring to the identities [60, eqs. (07.34.03.0271.01), (07.35.03.0001.01), (07.35.26.0003.01)], the Logarithm and the fraction terms can be transformed into Meijer-G functions as follows

$$\log(1+\varpi\gamma) = G_{2,2}^{1,2}\left(\varpi\gamma \;\middle|\; \begin{array}{c}1,1\\1,0\end{array}\right), \quad (63)$$

$$(\mathcal{A}_1\gamma + \beta_R)^{-(s+m_R)} = \frac{1}{\beta_R^{m_R+s}\Gamma(m_R+s)}$$
$$\times G_{1,1}^{1,1}\left(\frac{\mathcal{A}_1}{\beta_R}\gamma \;\middle|\; \begin{array}{c}1-s-m_R\\0\end{array}\right), \quad (64)$$

Using [60, eq. (07.34.16.0002.01)], we invert the argument of the Meijer-G function as follows:

$$G_{p+q+\alpha_2 p, \alpha_2 p}^{0, p+q+\alpha_2 p}\left(\mathcal{A}_4 \gamma^{-\frac{\alpha_2 p}{r}} \;\middle|\; \begin{array}{c}\kappa_1\\\kappa_2\end{array}\right)$$
$$= G_{\alpha_2 p, p+q+\alpha_2 p}^{p+q+\alpha_2 p, 0}\left(\frac{\gamma^{\frac{\alpha_2 p}{r}}}{\mathcal{A}_4} \;\middle|\; \begin{array}{c}[1]_{\text{length}(\kappa_1)}-\kappa_1\\{[1]_{\text{length}(\kappa_2)}-\kappa_2}\end{array}\right), \quad (65)$$

After this transformation, the term $\mathcal{C}_5$ consists of three Meijer-G functions. The next step is to expand these three functions using the general definition of the line integral in the complex plane [60, eq. (07.34.02.0001.01)] as follows

$$G_{1,1}^{1,1}\left(\frac{\mathcal{A}_1}{\beta_R}\gamma \;\middle|\; \begin{array}{c}1-s-m_R\\0\end{array}\right)$$
$$= \frac{1}{2\pi i}\int_{\mathcal{L}_1} \Gamma(-s_1)\Gamma(s+m_R+s_1) \times \left(\frac{\mathcal{A}_1}{\beta_R}\right) \gamma_1^{s_1} ds_1, \quad (66)$$

$$G_{2,2}^{1,2}\left(\varpi\gamma \;\middle|\; \begin{array}{c}1,1\\1,0\end{array}\right)$$
$$= \frac{1}{2\pi i}\int_{\mathcal{L}_2} \frac{\Gamma(1-s_2)\Gamma^2(s_2)}{\Gamma(s_2+1)} \varpi^{s_2} \gamma^{s_2} ds_2, \quad (67)$$

$$G_{\alpha_2 p, p+q+\alpha_2 p}^{p+q+\alpha_2 p, 0}\left(\frac{\gamma^{\frac{\alpha_2 p}{r}}}{\mathcal{A}_4} \;\middle|\; \begin{array}{c}[1]_{\text{length}(\kappa_1)}-\kappa_1\\{[1]_{\text{length}(\kappa_2)}-\kappa_2}\end{array}\right)$$
$$= \frac{1}{2\pi i}\int_{\mathcal{L}_3} \frac{\prod_{j=1}^{p+q+\alpha_2 p}\Gamma(1-\kappa_{2,j}-s_3)}{\prod_{j=1}^{\alpha_2 p}\Gamma(1-\kappa_{1,j}-s_3)} \mathcal{A}_4^{-s_3}\gamma^{\frac{\alpha_2 p s_3}{r}} ds_3, \quad (68)$$

The term $\mathcal{C}_5$ can be given by Eq. (70), as shown at the top of this page. Using [58, eq. (3.351.3)], the integral $\mathcal{I}$ is given by

$$\mathcal{I} = \Gamma\left(l+s_1+s_2+\frac{\alpha_2 p}{r}s_3+1\right)\mathcal{A}_1^{-(l+s_1+s_2+\frac{\alpha_2 p}{r}s_3+1)}, \quad (69)$$

Finally, using the identities [65, eqs. (28), (29a), (29b), (30)], the final result can be expressed in terms of the trivariate Fox-H function.

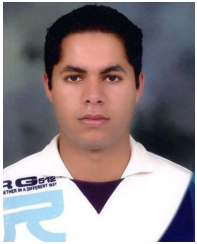

**Elyes Balti** (S'16) received the B.S. degree in electrical engineering from the Higher School of Communications of Tunis (Sup'Com), Tunisia, in 2013. He joined the University of Idaho, Moscow, Idaho, in 2016, where he is currently working toward the M.S. degree in electrical engineering. His research focuses on millimeter waves, optical wireless communications, fifth-generation cellular networks, and MIMO communication systems.

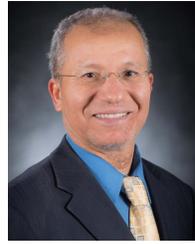

**Mohsen Guizani** (S'85–M'89–SM'99–F'09) received the B.S. (with distinction) and M.S. degrees in electrical engineering, and the M.S. and Ph.D. degrees in computer engineering from Syracuse University, Syracuse, NY, USA, in 1984, 1986, 1987, and 1990, respectively. He is currently a Professor and the ECE Department Chair at the University of Idaho, USA. Previously, he served as the Associate Vice President of Graduate Studies, Qatar University, Chair of the Computer Science Department, Western Michigan University, and Chair of the Computer Science Department, University of West Florida. He also served in academic positions at the University of Missouri-Kansas City, University of Colorado-Boulder, Syracuse University, and Kuwait University. His research interests include wireless communications and mobile computing, computer networks, mobile cloud computing, security, and smart grid. He is the author of 9 books and more than 450 publications in refereed journals and conferences. He currently serves on the editorial boards of several international technical journals and is the Founder and Editor-in-Chief of *Wireless Communications and Mobile Computing* (Wiley). He has guest edited a number of special issues in IEEE journals and magazines. He also served as a member, chair, and general chair of a number of international conferences. He received the teaching award multiple times from different institutions as well as the Best Research Award from three institutions. He was the Chair of the IEEE Communications Society Wireless Technical Committee and the Chair of the TAOS Technical Committee. He served as the IEEE Computer Society Distinguished Speaker from 2003 to 2005. He is a Senior Member of ACM.